\documentclass[a4paper,11pt]{article}
\pdfoutput=1
\usepackage{amssymb,amsmath,amsfonts,makeidx,placeins,pbox,multirow}
\usepackage{graphicx,rotate,subcaption,color,slashed,cite,caption,epstopdf,verbatim,url}
\usepackage{longtable,tabu}
\usepackage{array}
\newcolumntype{P}[1]{>{\centering\arraybackslash}p{#1}}
\usepackage[colorlinks=true,
            linkcolor=magenta,
            urlcolor=blue,
            citecolor=blue]{hyperref}
\usepackage[utf8x]{inputenc}

\numberwithin{equation}{section}

\evensidemargin=\oddsidemargin
\begin{document}

\begin{center}

{\Large \bf Impact of Yukawa-like dimension-5 operators on\\\vspace{0.12cm} the Georgi-Machacek model} \\
\vspace*{0.5cm} {\sf Avik
  Banerjee~\footnote{avik.banerjeesinp@saha.ac.in}, ~Gautam
  Bhattacharyya~\footnote{gautam.bhattacharyya@saha.ac.in}, ~Nilanjana Kumar~\footnote{nilanjana.kumar@gmail.com}} \\
\vspace{10pt} {\small } {\em Saha Institute of Nuclear
    Physics, HBNI, 1/AF Bidhan Nagar, Kolkata 700064, India}
\normalsize
\end{center}


\begin{abstract}

We study the effects of including Yukawa-like dimension-5 operators in the Georgi-Machacek model where the Standard Model is augmented with triplet scalars. We focus only on the charged Higgs sector and investigate the constraints arising from radiative $B$-meson decays, neutral $B$-meson mixing and precision measurement of $Zb\bar{b}$ vertex. We observe that the inclusion of the dimension-5 operators causes substantial alteration of the limits on the charged Higgs masses and the vacuum expectation value of the triplets, derived otherwise using only the dimension-4 operators.
\end{abstract}


\bigskip


\section{Introduction}
\label{intro}

The Standard Model (SM) Higgs-like scalar boson, discovered at the Large hadron collider (LHC), has not yet provided a full understanding of the dynamics of electroweak symmetry breaking. Many motivated Beyond the SM (BSM) scenarios postulate the presence of additional scalars (e.g.\ $\rm SU(2)$ singlet/ doublet/ triplet). Exploration of these scalars through direct searches at colliders or through their indirect contributions in various observables is an ongoing exercise. In the absence of any new physics signals at the LHC so far, an effective field theory approach with higher dimensional operators has been employed in different contexts \cite{Buchmuller:1985jz,Hagiwara:1993ck,Giudice:2007fh,Grzadkowski:2010es,Ellis:2014dva}. Construction of effective theories in the singlet and two Higgs doublet extended scenarios has received considerable attention \cite{Chala:2017sjk,DiazCruz:2001tn,Kikuta:2011ew,Kikuta:2015pya,Crivellin:2016ihg,Karmakar:2017yek}. In this paper we follow a bottom-up approach in studying the triplet-extended Higgs models in the effective field theoretic framework keeping Yukawa-type operators up to dimension-5. 

For illustration, we choose a particular type of Higgs triplet
scenario, known as the Georgi-Machacek (GM) model, which protects
custodial symmetry at tree level
\cite{Georgi:1985nv,Chanowitz:1985ug,Gunion:1989ci,Aoki:2007ah,Chiang:2012cn,Hartling:2014zca,Hartling:2014aga,Chiang:2014hia,Chiang:2014bia,Logan:2015xpa,Chiang:2015amq,Chang:2017niy,Degrande:2017naf,Blasi:2017xmc,Biswas:2018jun,Keeshan:2018ypw}. In
addition to the renormalizable Yukawa terms involving the SM doublet
Higgs, we include dimension-5 operators in the quark sector
(especially, the third generation) employing the scalar triplet as
well. Note that in two Higgs doublet models the only dimension-5
operator is the Weinberg operator which appears in the leptonic sector
generating neutrino Majorana masses \cite{Weinberg:1979sa}. On the
other hand, the scalar triplets can couple to the lepton doublets
through a renormalizable dimension-4 operator leading to type-II
see-saw.  Further inclusion of dimension-5 terms involving the
triplets, together with the usual Weinberg operator, would invariably
add new twists to neutrino phenomenology.  In this paper, however, our
main focus is to explore indirect constraints on the charged Higgs
sector in the GM model by admitting dimension-5 operators that involve
the scalar triplets and the heavy third generation quarks.

We note that the presence of higher dimensional Yukawa-like operators
is, in fact, a generic feature of a large class of BSM theories. They
can arise, for example, when heavy vector-like fermions are integrated
out \cite{Leskow:2014kga,Chala:2018opy}. A broad class of composite
Higgs models (e.g.\ little Higgs models, gauge-Higgs unification
models, etc.) contains such heavy fermions
\cite{Contino:2010rs,Csaki:2016kln,Panico:2015jxa}.  Instead of
appealing to any specific BSM theory, we construct a toy scenario with
independent dimension-5 operators involving the quarks with unknown
$\mathcal{O}(1)$ coefficients, and constrain the parameter space using
flavor and electroweak observables, more specifically, $B\rightarrow
X_s\gamma$, $B^0_s-\bar{B}^0_s$ mixing and $Zb\bar{b}$
vertex. Depending on the relative sign and magnitude of the
coefficients of these dimension-5 operators, the constraints on the
charged Higgs masses can be substantially modified from the existing
limits obtained employing the usual dimension-4 operators only.

\section{The setup and the relevant operators}
\label{model}

We outline here the salient features of the GM model and construct the dimension-5 operators relevant for our discussions. The GM model contains two Higgs triplets with hypercharge $Y=0$ and $Y=1$, in addition to the SM Higgs doublet with $Y=1/2$. In this scenario, the scalar potential preserves custodial symmetry at tree level without keeping the triplet scalars necessarily inert. This is possible because the Higgs doublet and the triplets can be embedded in the following bi-doublet and bi-triplet representations under $\rm SU(2)_L\times SU(2)_R$, respectively, as 
\begin{equation}
\label{model_eq1}
\Phi=\left(\begin{tabular}{cc}
$\phi^{0*}$ &  $\phi^+$\\
$-\phi^-$ & $\phi^{0}$
\end{tabular}\right),~~\Delta=\left(\begin{tabular}{ccc}
$\chi^{0*}$ & $\xi^+$ & $\chi^{++}$\\
$-\chi^{-}$ & $\xi^0$ & $\chi^{+}$\\
$\chi^{--}$ & $-\xi^-$ & $\chi^{0}$
\end{tabular}\right).
\end{equation}
Details of the scalar potential and its minimization can be found in \cite{Chiang:2013rua,Hartling:2014zca}. We assume that $\langle\phi^0\rangle=v_d/\sqrt{2}$, $\langle\chi^0\rangle=\langle\xi^0\rangle=v_t$, and $v^2=v_d^2+8v_t^2\simeq (246~\textrm{GeV})^2$. Using the standard notation in literature, we define
\begin{equation}
\label{model_eq2}
\tan\beta\equiv{2\sqrt{2}v_t\over v_d}~.
\end{equation}
Under the custodial $\rm SU(2)_V$ one constructs the 5-plet $(H_5^{\pm\pm},H_5^{\pm},H_5^0)$, triplet $(H_3^{\pm},H_3^0)$ and two singlet $(h, H)$ scalars. With a slight perversion of notation, we shall represent the mass eigenstates after diagonalization using the same symbols. Their expressions in terms of the original fields in Eq.~\eqref{model_eq1} are given in \cite{Chiang:2013rua,Hartling:2014zca}.

We now focus on the couplings of the scalars with the quarks. Since the Yukawa couplings explicitly break custodial symmetry due to different hypercharge assignments for the left- and right-handed fermions, instead of expressing the scalars as bi-doublets and bi-triplets, we rather use $2_{1/2}$, $3_0$ and $3_1$ representations under $\rm SU(2)_L\times U(1)_Y$, as follows : 
\begin{equation}
\label{model_eq3}
\phi=\left(\begin{tabular}{c}
$\phi^+$\\ 
$\phi^0$ 
\end{tabular}\right),~~
\xi=\left(\begin{tabular}{cc}
$\xi^0/\sqrt{2}$ & $-\xi^+$\\ 
$-\xi^-$ & $-\xi^0/\sqrt{2}$ 
\end{tabular}\right),~~
\chi=\left(\begin{tabular}{cc}
$\chi^+/\sqrt{2}$ & $-\chi^{++}$\\ 
$\chi^0$ & $-\chi^+/\sqrt{2}$ 
\end{tabular}\right).
\end{equation}
The usual dimension-4 Yukawa Lagrangian is given by
\begin{equation}
\label{model_eq4}
-\mathcal{L}^{(4)}_{\textrm{Yuk}}=y^u_{ij}\bar{Q}_{Li}\phi^cu_{Rj}+y^d_{ij}\bar{Q}_{Li}\phi d_{Rj}+\textrm{h.c.}~,
\end{equation}
which is not expected to contain the triplets $\xi$ and $\chi$ for group theoretic reason.
\begin{table}[t]
	\begin{longtable}{ccccc}
		\hline 
		\rule[-2ex]{0pt}{5.5ex} Vertices &  & Feynman Rules &\\ 
		\hline  
		\hline
		\\
		
		\rule[-2ex]{0pt}{5.5ex} $h\bar{f}f$ & & $-i{m_f\over v}\left[{c_\alpha\over c_\beta}+s_\alpha\left({c^f_5\over \sqrt{3}}\pm{d^f_5\over \sqrt{6}}\right){v\over \Lambda}\right]$ & \\\\
		
		\rule[-2ex]{0pt}{5.5ex} $H\bar{f}f$ & & $-i{m_f\over v}\left[-{s_\alpha\over c_\beta}+c_\alpha\left({c^f_5\over \sqrt{3}}\pm{d^f_5\over \sqrt{6}}\right){v\over \Lambda}\right]$ & \\\\
		
		\rule[-2ex]{0pt}{5.5ex} $H^0_3\bar{f}f$ & & $\pm\gamma_5{m_f\over v}\left[t_\beta-{c^f_5\over \sqrt{2}}{1\over c_\beta}{v\over \Lambda}\right]$ & \\\\
		
		\rule[-2ex]{0pt}{5.5ex} $H^0_5\bar{f}f$ & & $-i{m_f\over v}\left[\left({c^f_5\over \sqrt{6}}\mp{d^f_5\over \sqrt{3}}\right){v\over\Lambda}\right]$ & \\\\
		
		\hline
		
		\rule[-2ex]{0pt}{5.5ex} $H^+_3\bar{u}d$ & & $-i{\sqrt{2}\over v}V_{ud}\left[\left(t_\beta-{1\over c_\beta}\left({c^u_5\over 2\sqrt{2}}+{d^u_5\over 2}\right){v\over \Lambda}\right)m_uP_L-\left(t_\beta-{1\over c_\beta}\left({c^d_5\over 2\sqrt{2}}-{d^d_5\over 2}\right){v\over \Lambda}\right)m_dP_R\right]$ & \\\\
		
		\rule[-2ex]{0pt}{5.5ex} $H^+_5\bar{u}d$ & & $i{\sqrt{2}\over v}V_{ud}\left[\left({c^u_5\over 2\sqrt{2}}-{d^u_5\over 2}\right){v\over \Lambda}m_uP_L-\left({c^d_5\over 2\sqrt{2}}+{d^d_5\over 2}\right){v\over \Lambda}m_dP_R\right]$ & \\\\
		
		\hline 
		\caption{\sf\it The couplings of the quarks with the physical scalars are listed. The relative $\pm$ sign appearing in Feynman rules refer to up/down quarks. Also $s_\alpha(c_\alpha)\equiv\sin\alpha(\cos\alpha)$, where $\alpha$ is the mixing angle between the neutral scalars and $t_\beta(c_\beta)\equiv\tan\beta(\cos\beta)$. For our calculations, $H_{3,5}^\pm$ couplings are relevant.} 
		\label{dim5yuk_tab1}
	\end{longtable}
\end{table}
The triplets, however, can couple to the quarks through dimension-5 operators, as
\begin{equation}
\label{model_eq5}
-\mathcal{L}^{(5)}_{\textrm{Yuk}}={c^u_5\over \Lambda}y^u_{ij}\bar{Q}_{Li}\chi^\dagger \phi u_{Rj}+{c^d_5\over \Lambda}y^d_{ij}\bar{Q}_{Li}\chi \phi^cd_{Rj}+{d^u_5\over \Lambda}y^u_{ij}\bar{Q}_{Li}\xi \phi^cu_{Rj}+{d^d_5\over \Lambda}y^d_{ij}\bar{Q}_{Li}\xi \phi d_{Rj}+\textrm{h.c.}~,
\end{equation}
where $\Lambda$ is the cut-off scale of the effective operators. Following the minimal flavor violation hypothesis the coefficients of dimension-5 operators are assumed to be aligned with the Yukawa couplings to avoid stringent constraints from flavor changing neutral current processes. The coefficients $c_5^{u,d}$ and $d_5^{u,d}$ are taken as $\mathcal{O}(1)$ real numbers, whose exact values and signs depend on the specific underlying UV models. Note that the real coefficients keep the 125 GeV Higgs boson as purely CP-even. In this paper we will treat these coefficients as free parameters. The couplings of the quarks with the physical scalars are displayed in Table~\ref{dim5yuk_tab1}. While the couplings of the 5-plet scalars arise only at the dimension-5 level, those involving the triplet and the singlet scalars originate from both dimension-4 and dimension-5 terms.

\section{Flavor and electroweak phenomenology}
\label{constraints}

The most significant constraints on the charged Higgs masses and couplings in this setup would come from the radiative $B$ decay, neutral $B$-meson mixing and the precision measurement of the $Zb\bar{b}$ vertex, which we discuss below.

\subsection{$B\rightarrow X_s\gamma$ decay}
\label{bsgamma}

The latest experimental world average for $\textrm{Br}^{\textrm{exp}}(B\rightarrow X_s\gamma)=(3.32\pm 0.16)\times 10^{-4}$ \cite{Amhis:2016xyh}, while $\textrm{Br}^{\textrm{SM}}(B\rightarrow X_s\gamma)=(3.36\pm 0.23)\times 10^{-4}$ \cite{Misiak:2015xwa}. The branching ratio receives large contributions from the charged Higgs couplings \emph{via} the Wilson coefficient $C^{\rm eff}_7$. The structure of the charged Higgs $(H_i^\pm)$ couplings with the quarks is given by
\begin{equation}
\mathcal{L}={\sqrt{2}\over v}V_{ud}H_i^+\bar{u}\left[A^i_um_uP_L-A^i_dm_dP_R\right]d+\textrm{h.c.}
\end{equation}
The new physics contributions to $C^{\rm eff}_{7,8}$ (dominated by the top quark mass) at the matching scale ($\sim 160$ GeV \cite{Misiak:2015xwa}) is given by
\begin{equation}
\delta C^{\textrm{eff}}_{7,8}=\sum_{i}\left[{(A^i_{t})^2\over 3}F^{(1)}_{7,8}(x_{i})-A^{i}_{t}A^i_{b} F^{(2)}_{7,8}(x_{i})\right],
\end{equation}
where $x_{i}\equiv m_t^2/m_i^2$. Here $i$ takes two values, $3$ and $5$, which correspond to the scalars $H_3^\pm$ and $H_5^\pm$. We restrict ourselves to leading order in new physics. The functions $F_{7,8}^{(1,2)}(x_{i})$ can be found in \cite{Ciuchini:1997xe}. The expressions for $A_{t,b}^i$ can be read off from Table~\ref{dim5yuk_tab1}. Following \cite{Misiak:2015xwa,Hu:2016gpe}, we have translated the limits on branching ratio to the following range
\begin{equation}
-0.063\le\delta C^{\textrm{eff}}_7+0.242~\delta C^{\textrm{eff}}_8\le 0.073~,
\end{equation}
where we have combined the theoretical and experimental uncertainties in quadrature.
With only the dimension-4 operators, the new physics part of the GM model always contributes destructively with the SM amplitude leading to a decrease in the overall branching ratio. The dimension-5 operators contribute constructively or destructively depending on the sign of the coefficients. However, since the numerical impact of dimension-5 operators is smaller than that of dimension-4 operators, the prediction for the overall branching ratio stays reduced compared to the SM expectation. 

\subsection{Neutral $B$-meson mixing}
\label{BBbar}

For the sake of demonstration, we show the charged Higgs contributions to the $B_s^0-\bar{B}_s^0$ mixing, as it provides slightly stronger constraints than $B_d^0-\bar{B}_d^0$ mixing. The primary reason for this is smaller uncertainties in $B_s$-system for both experimental measurements and the SM predictions \cite{Amhis:2016xyh}. The measured value of the mass splitting and its SM prediction in $B_s$-system are given by\cite{Amhis:2016xyh}
\begin{equation}
\Delta m^{\textrm{exp}}_{B_s}=(17.757\pm 0.021)~\textrm{ps}^{-1},~~~\Delta m^{\textrm{SM}}_{B_s}=(18.3\pm 2.7)~\rm ps^{-1}.
\end{equation}
We obtain the total contributions to the mass splitting from the $W^\pm$ bosons, the Goldstones and the charged Higgs bosons ($H_3^\pm, H_5^\pm$), through the box graphs using standard notations as
\begin{equation}
\Delta m_{B_s}={G_F^2m_t^2\over 24\pi^2}(V^*_{ts}V_{tb})^2f_{B_s}^2B_{B_s}m_{B_s}\eta_BI_{\textrm{tot}}(x_W,x_i,x_j)~,
\end{equation}
where $x_W=m_t^2/M_W^2$, and
\begin{eqnarray}
I_{\textrm{tot}}=I_{WW}(x_W)+\sum_{i,j}(A^i_{t})^2(A^j_{t})^2I_{H_iH_j}(x_{i},x_{j})
+2\sum_{i}(A^i_{t})^2I_{WH_i}(x_W,x_{i})~.
\end{eqnarray}
The explicit expressions for the Inami-Lim functions $I_{WW}$, $I_{WH_i}$ and $I_{H_iH_i}$ can be found in \cite{Mahmoudi:2009zx,Abbott:1979dt}, while we have calculated $I_{H_iH_j}$ (with $i\ne j$), given by
\begin{eqnarray}
I_{H_iH_j}= x_i x_j\left[{1\over (1-x_i)(1-x_j)}+{\log x_i\over (x_i-x_j)(1-x_i)^2}+{\log x_j\over (x_j-x_i)(1-x_j)^2}\right].
\end{eqnarray}
Normalizing $\Delta m_{B_s}$ with respect to its SM prediction, we obtain the following range at $2\sigma$
\begin{equation}
0.675 \le{\Delta m_{B_s}\over \Delta m^{\textrm{SM}}_{B_s}}={I_{\textrm{tot}}(x_W,x_i,x_j)\over I_{WW}(x_W)}\le 1.265~. 
\end{equation}
Unlike in the $B\rightarrow X_s\gamma$ case, the dimension-4 new physics contributions add up constructively with the SM part in neutral meson mixing. However, the dimension-5 contributions depend on the sign of $c_5^t$ and $d^t_5$. 

\subsection{$Zb\bar{b}$ vertex}
\label{Zbb}

One of the most precisely measured electroweak observables is the $Z\rightarrow b\bar{b}$ branching ratio
\begin{equation}
R_b={\Gamma(Z\rightarrow b\bar{b})\over \Gamma(Z\rightarrow \rm hadrons)}~,
\end{equation}
where $R_b^{\rm exp}=0.21629\pm0.00066$ \cite{ALEPH:2005ab} and $R_b^{\rm SM}=0.21581\pm0.00011$ \cite{Haller:2018nnx}. The modifications in $R_b$ due to the charged Higgs contributions at one loop is given by
\begin{eqnarray}
\delta R_b\simeq-0.7785~\delta g^L_\textrm{new}~.
\end{eqnarray}
Here $\delta g^L_\textrm{new}$ is the modification in the $Zb_L\bar{b}_L$ coupling, calculated from a combination of triangle graphs where $H^\pm_{3,5}$ and the charged Goldstones float inside the loop. Their explicit expressions can be found in \cite{Haber:1999zh,Logan:1999if}. A new type of triangle graph, induced by the dimension-5 operators, nevertheless arises in our context. This involves the set $\{H^\pm_5,W^\mp,t\}$ inside the loop. Its contribution is given by
\begin{equation}
\delta g^L_{\textrm{new}}(H^\pm_5,W^\mp,t)=-{g^2\over 16\pi^2}|V_{tb}|^2\left({c^t_5\over 2\sqrt{2}}-{d^t_5\over 2}\right){v\over \Lambda}s_\beta m_t^2C_0(m_t,M_W,m_5)~,
\end{equation}
where $C_0(m_t,M_W,m_5)$ is the usual Passarino-Veltman function \cite{Passarino:1978jh} and $g$ denotes the $\rm SU(2)_L$ gauge coupling. This new graph provides a numerically significant interference with the other contributions. We constrain the new physics parameter space by squeezing it in the following $2\sigma$ range
\begin{equation}
-0.00086\le \delta R_b\le 0.00182~.
\end{equation}  

\subsection{Results}
\label{resuls}

Before discussing our results, it is worthwhile to recall the existing constraints in the GM model that guided us choose our benchmark values. The oblique S and T parameters do not sense the dimension-5 Yukawa-like operators at one loop. Nevertheless, they constrain the vacuum expectation value (vev) of the triplet scalars from the dimension-4 operators in the gauge sector \cite{Chiang:2013rua,Kanemura:2013mc,Hartling:2014aga}. The $125$ GeV Higgs boson production and decay are of course affected by the dimension-5 operators. This, however, involves the neutral Higgs mixing angle ($\alpha$), which can be tuned to restrict the severity of the contribution \cite{Logan:2010en,Belanger:2013xza,Kanemura:2013mc,Chiang:2015kka,Chiang:2015amq,Li:2017daq,Chiang:2018cgb,Das:2018vkv}. Non-observation from direct searches also restrict the masses of $H^\pm$ and $H^{\pm\pm}$, though the strategies involve several assumptions \cite{Li:2017daq,Cen:2018okf,Aad:2015nfa,Sirunyan:2017sbn,Chiang:2012dk,Englert:2013wga,Chiang:2015rva,Sun:2017mue,CMS-PAS-HIG-16-036,Aaboud:2017qph,Aaboud:2018qcu}.

\begin{figure}[t]
	\centering
	\begin{subfigure}[t]{0.415\textwidth}
		\centering
		\includegraphics[width=\linewidth]{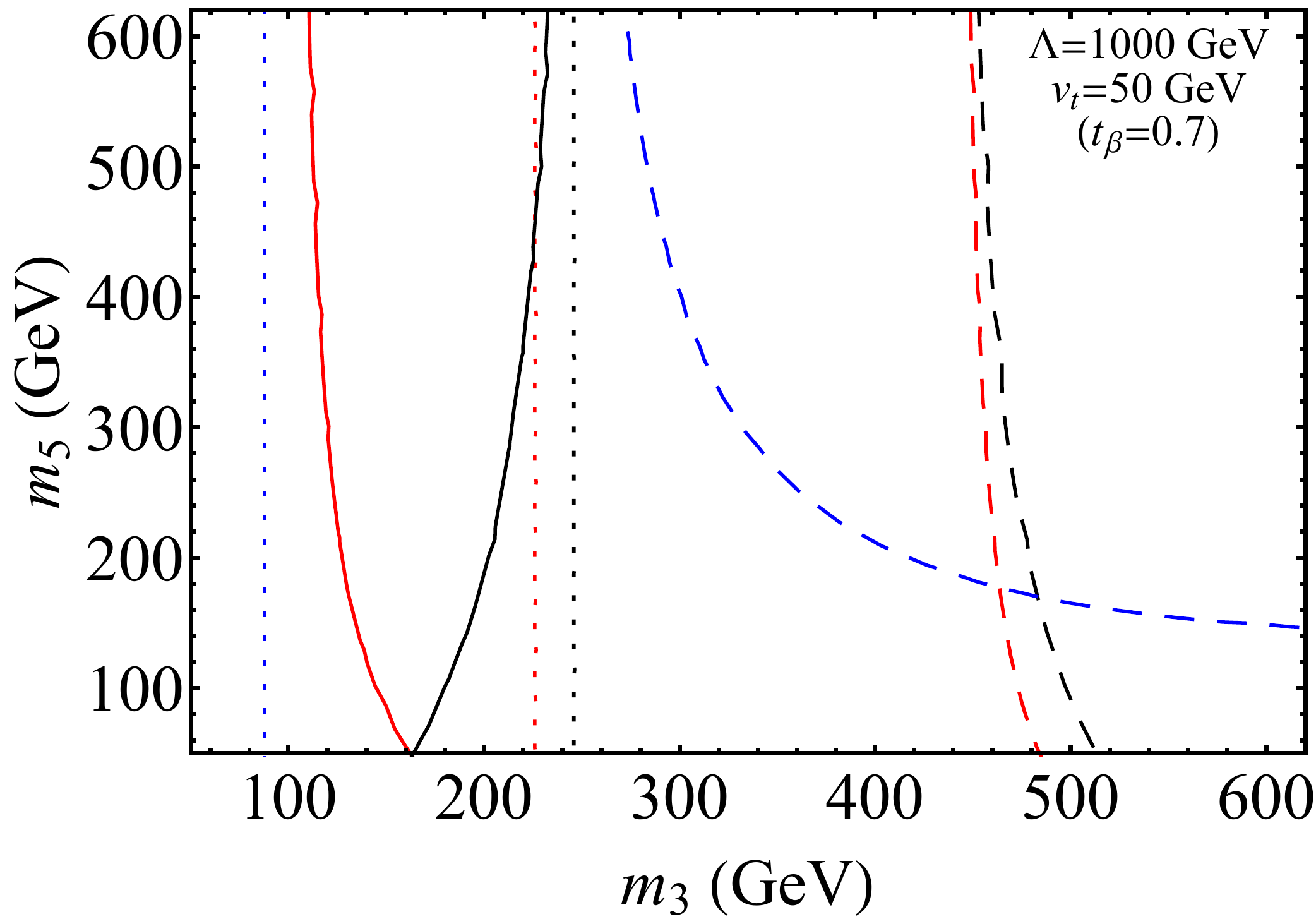}
	\end{subfigure}
	\begin{subfigure}[t]{0.55\textwidth}
		\centering
		\includegraphics[width=\linewidth]{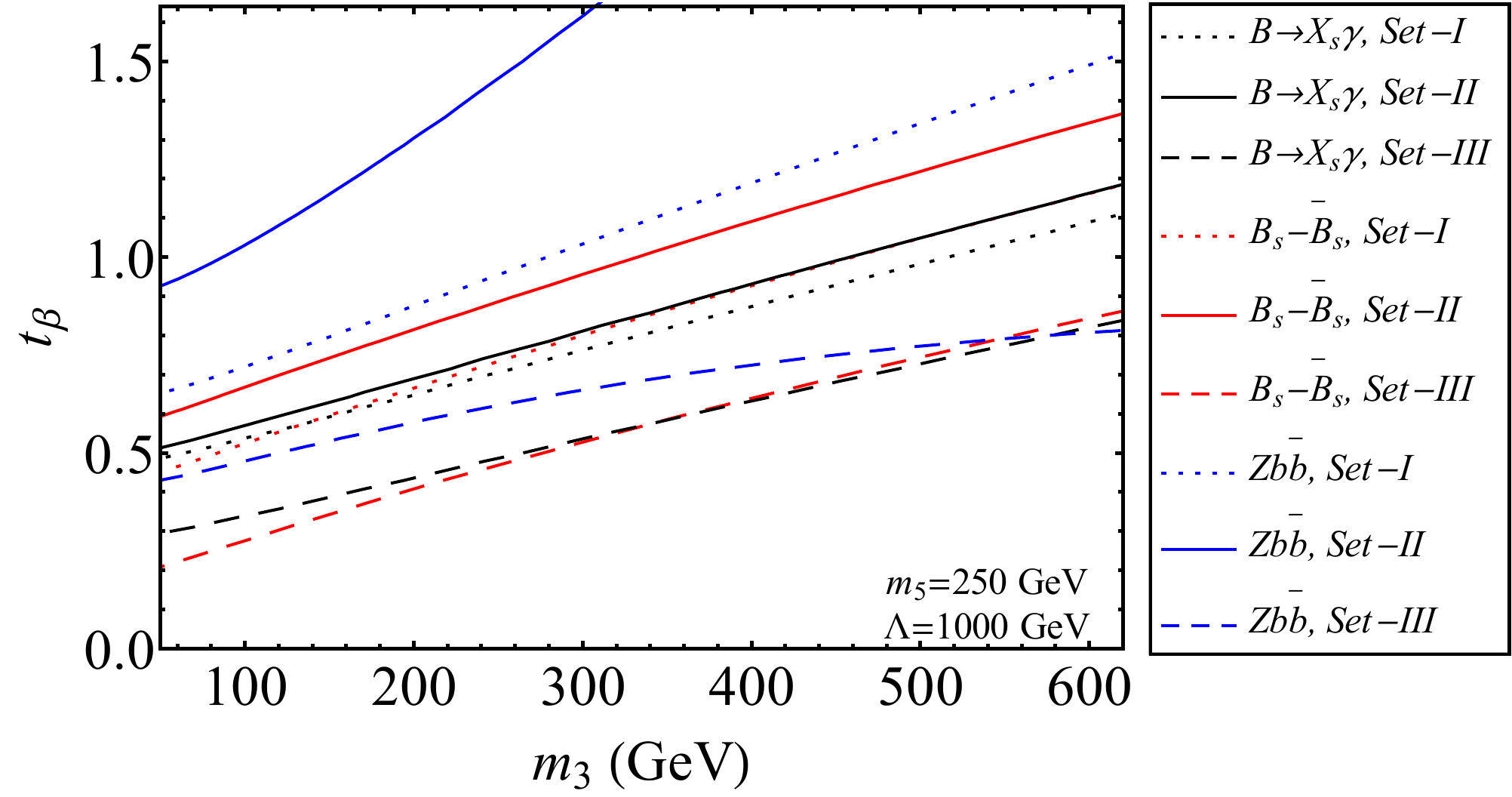}
	\end{subfigure}
	\caption{\small\it Exclusion limits on the charged Higgs masses and the triplet vev from $B\rightarrow X_s\gamma$, $B_s^0-\bar{B}_s^0$ mixing and $Zb\bar{b}$ vertex for the different sets of benchmark parameters (shown in Table~\ref{dim5yuk_tab2}). In the left panel, regions on the left of each curve are disfavored at $2\sigma$, while in the right panel the regions above each curve are disfavored at $2\sigma$. We have fixed $\Lambda=1$ TeV, and have taken $v_t=50$ GeV (left panel) and $m_5=250$ GeV (right panel).}
	\label{fig_1}
\end{figure}
\begin{figure}[t]
	\centering
		\includegraphics[width=0.45\textwidth]{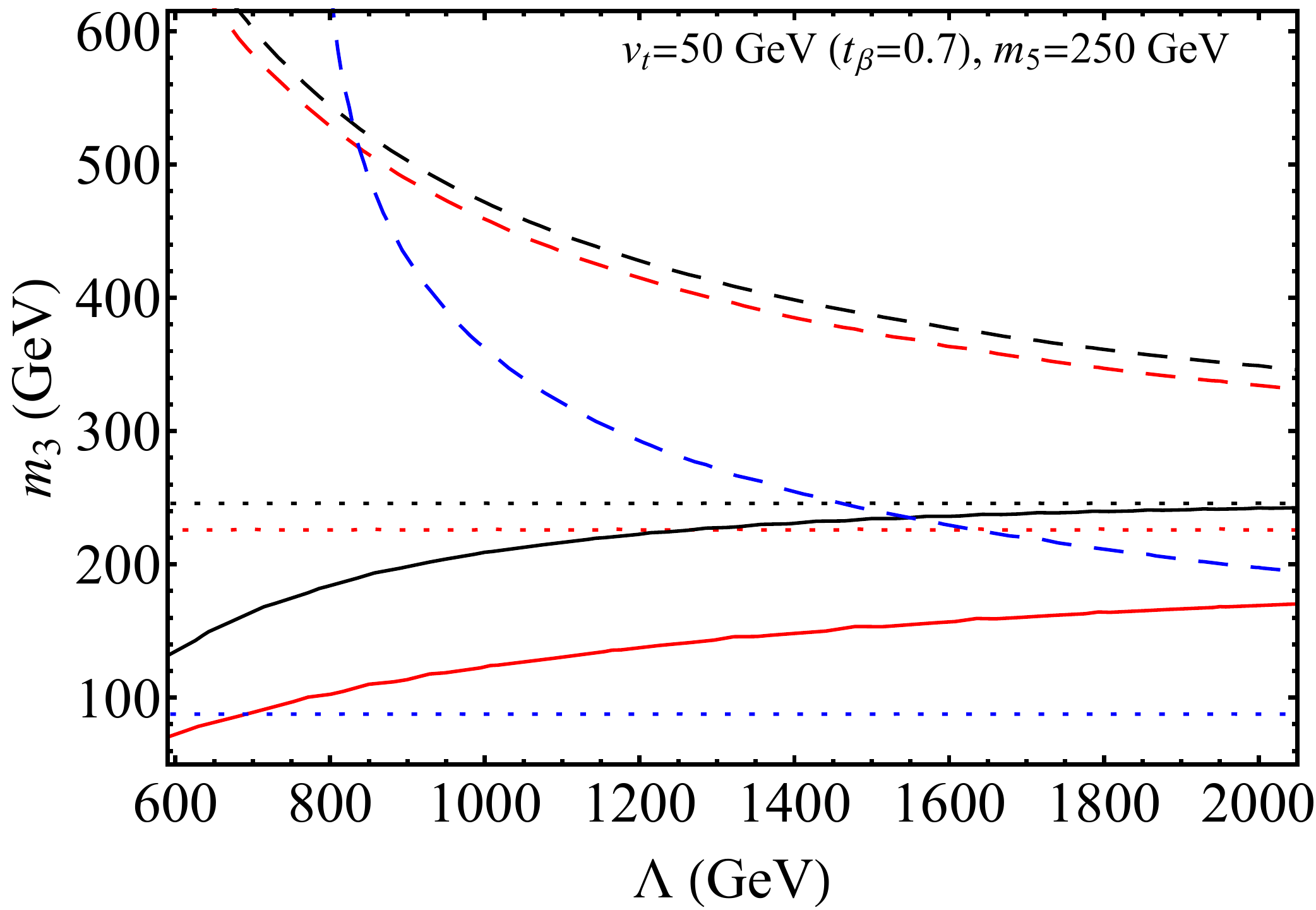}
	\caption{\small\it The cut-off scale $\Lambda$ is varied here fixing $v_t=50$ GeV and $m_5=250$ GeV. All the regions below each curve are disfavored at $2\sigma$. The implications of different lines are as in Fig.~\ref{fig_1}.}
	\label{fig_2}
\end{figure}

\begin{table}[]
	\begin{longtable}{cccccccccccccccccc}
		\hline 
		Set&\multicolumn{4}{c}{Benchmark parameters}& & \multicolumn{8}{c}{$2\sigma$ Lower limits on $m_3$, $m_5$ (in GeV)}\\
		& & & & & &\multicolumn{2}{c}{$B\rightarrow X_s\gamma$} & & \multicolumn{2}{c}{$B_s^0-\bar{B}_s^0$} & & \multicolumn{2}{c}{$Zb\bar{b}$}& \\
		\hline
		&$c^t_5$ & $d^t_5$ & $c^b_5$ & $d^b_5$& & $m_3$ & $m_5$ & & $m_3$ & $m_5$ & & $m_3$ & $m_5$ &\\
		\hline  
		\hline		
		
		&\multicolumn{13}{c}{$v_t=50$ GeV ($\tan\beta=0.70$)}\\
		\hline
		
		$\rm I$ & 0.0 & 0.0 & 0.0 & 0.0 & & $250$ & $-$ & & $225$ & $-$ & & $85$ & $-$ &\\
		
		$\rm II$ & -0.5 & 1.5 & 0.5 & 1.5 & & $230$ & $-$ & & $110$ & $-$ & & $-$ & $-$ &\\
		
		$\rm III$ & -0.5 & -1.5 & 0.5 & 1.5 & & $455$ & $-$ & & $450$ & $-$ & & $260$ & $145$ &\\
		\hline
		&\multicolumn{13}{c}{$v_t=40$ GeV ($\tan\beta=0.52$)}\\
				\hline
				$\rm I$ & 0.0 & 0.0 & 0.0 & 0.0 & & $80$ & $-$ & & $100$ & $-$ & & $-$ & $-$ &\\
				
				$\rm II$ & -0.5 & 1.5 & 0.5 & 1.5 & & $75$ & $-$ & & $-$ & $-$ & & $-$ & $-$ &\\
				
				$\rm III$ & -0.5 & -1.5 & 0.5 & 1.5 & & $270$ & $-$ & & $285$ & $-$ & & $115$ & $50$ &\\

		\hline
		\caption{\sf\it Modified lower limits on $m_3$ and $m_5$ in the presence of dimension-5 operators from three observables for different sets of input parameters. Clearly, the Set-I input values correspond to the purely dimension-4 case.} 
		\label{dim5yuk_tab2}
	\end{longtable}
\end{table}
We now discuss our observations. The left panel of Fig.~\ref{fig_1} shows the constraints in the plane of the charged Higgs masses coming from the triplet and 5-plet scalars ($m_3$ and $m_5$) for a few benchmark points chosen for illustration, while the right panel displays the same in the $\tan\beta-m_3$ plane. We draw attention to the substantial contributions from the dimension-5 operators when compared to the situation containing only dimension-4 terms. The sign and magnitude of the coefficients of the new operators play a crucial r\^ole here. In Table~\ref{dim5yuk_tab2}, we present some conservative lower limits on $m_3$, and in some cases also on $m_5$, for the above mentioned benchmark values of the parameters including the triplet vev. Larger the triplet vev, stronger is the constraint, as expected. Note that the constraints on $m_5$ arise only when we consider the dimension-5 operators. Among the three sets of observables, only $Zb\bar{b}$ offers reasonable constraints on $m_5$. The situation with $B\rightarrow X_s\gamma$ has become a little tricky over the last few years \cite{Amhis:2016xyh,Amhis:2014hma}. The experimentally measured central value and the SM prediction have moved in such a way that there exists more space to squeeze our parameters now than a few years ago. Consequently, the limits on the charged Higgs masses in the GM model from $B\rightarrow X_s\gamma$ are not as stringent as before \cite{Hartling:2014aga}. Throughout our analysis we have kept the cut-off scale $\Lambda=1$ TeV, except in Fig.~\ref{fig_2} where we plotted $m_3$ against $\Lambda$, fixing other parameters. We emphasize that the constraints we derived on $m_3$, $m_5$ and $\tan\beta$, \textit{albeit} depending on the benchmark values, are both complementary as well as competitive with those obtained from oblique electroweak parameters and direct searches.

\section{Conclusions and outlook}
\label{conclusion}

Two points are worth noting in our analysis. First, we have constructed the dimension-5 Yukawa-like effective operators from a bottom-up phenomenological approach in the Georgi-Machacek framework keeping their UV origin unspecified. Secondly, we demonstrate that for reasonable values of the input parameters these operators significantly modify the limits on the charged Higgs masses. In this context, the dimension-5 operators provide a new handle to constrain the mass of the 5-plet charged Higgs ($H_5^\pm$). All we emphasize is that the limits on the charged Higgs masses derived previously are not infallible once we admit higher dimensional operators. Therefore, while devising the search strategies one should not be biased by the previously existing limits.

A natural extension of our study would be to construct similar operators in the leptonic sector, and importantly a new one, given by $l^T_LCi\tau_2\chi\xi l$, together with the standard Weinberg operator $(l^T_LCi\tau_2\phi)(\phi^T i\tau_2l)$. A further extension can be envisaged by constructing the full set of higher dimensional operators in both Yukawa and gauge sectors in an effective theory framework at the expense of introducing more $\mathcal{O}(1)$ parameters which would affect a large pool of observables.

\begin{small}
\section*{Acknowledgments}

The authors acknowledge support from the Indo-French Center for Promotion of Advanced Research (IFCPAR/CEFIPRA, Project No. 5404-2). AB acknowledges support from the Department of Atomic Energy, Government of India and thanks LPT, Orsay and \'Ecole Polytechnique, Palaiseau for the hospitality during the initial stages of the work. GB acknowledges support of the J.C. Bose National Fellowship from the Department of Science and Technology, Government of India (SERB Grant No. SB/S2/JCB-062/2016).

\end{small}



\bibliographystyle{JHEP}
\bibliography{triplet_flavor}

\end{document}